\documentclass[12pt]{iopart}
\usepackage{amsfonts,amssymb}
\usepackage{graphicx}

\newcommand{\ind}[2]{^{#1}_{\mbox{\scriptsize #2}}}
\newcommand{\al}[2]{\alpha\ind{#1}{#2}}
\newcommand{\Nc}{N_{\mbox{\scriptsize c}}}
\newcommand{\nf}{n_{\mbox{\scriptsize f}}}
\newcommand{\rd}{\rho_{\mbox{\tiny{D}}}}
\newcommand{\rp}[1]{\rho\ind{#1}{pert}}
\newcommand{\mpi}{m_{\pi}}
\newcommand{\epem}{e^{+}e^{-}}
\newcommand{\pc}{\,\%}

\begin{document}

\title[]{A~novel~integral~representation~for~the~Adler~function}

\author{A V Nesterenko$^{1,2}$ and J Papavassiliou$^{1}$}


\address{$^{1}$ Departamento de F\'\i sica Te\'orica
and IFIC, Centro Mixto, \\ Universidad de Valencia--CSIC,
E-46100, Burjassot, Valencia, Spain}

\address{$^{2}$ Bogoliubov Laboratory of Theoretical Physics,
Joint Institute for Nuclear Research, \\
Dubna, 141980, Russian Federation}

\ead{Joannis.Papavassiliou@uv.es}


\begin{abstract}
New integral representations for the Adler $D$--function and the
$R$--ratio of the electron--positron annihilation into hadrons are
derived in the general framework of the analytic approach to QCD.
These representations capture the nonperturbative information encoded
in the dispersion relation for the $D$--function, the effects due to
the interrelation between spacelike and timelike domains, and the
effects due to the nonvanishing pion mass. The latter plays a crucial
role in this analysis, forcing the Adler function to vanish in the
infrared limit. Within the developed approach the $D$--function is
calculated by employing its perturbative approximation as the only
additional input. The obtained result is found to be in reasonable
agreement with the experimental prediction for the Adler function in
the entire range of momenta  $0 \le Q^2 < \infty$.
\end{abstract}

\submitto{\jpg}


\maketitle

\section{Introduction}

     The hadronic vacuum polarization function $\Pi(q^2)$ and the
Adler function $D(Q^2)$ are physical observables of central
importance in various fields of elementary particle physics. In
particular, these quantities are essential for the analysis of strong
interaction processes such as electron--positron annihilation into
hadrons~\cite{Feynman,Adler,Georgi,Bjorken} and $\tau$~lepton
decay~\cite{Tau,BNP}. Furthermore, the Adler function plays an
important role when confronting precise experimental measurements of
some electroweak observables  with their theoretical predictions,
giving rise to decisive tests of the Standard Model and furnishing
stringent  constraints on possible new  physics beyond it.
Specifically, due to the hadronic excitations in the photon vacuum
polarization, the muon anomalous magnetic
moment~\cite{MuonExp,MuonTheo} and the shift of the electromagnetic
fine structure constant~\cite{RCQEDExp,RCQEDTheo} receive appreciable
contributions from the strong interactions. The latter constitute in
fact the main source of theoretical uncertainty when computing these
observables, since the physical energy scales involved are such that
the nonperturbative effects begin to be relevant.

     To date, there is no systematic method for calculating the Adler
function in the whole range of energies $0 \le Q^2 < \infty$.
Nonetheless, in the asymptotic ultraviolet region $D(Q^2)$ may be
approximated by the power series in the strong running
coupling~$\al{}{s}(Q^2)$ by making use of perturbation theory.
However, unphysical singularities  of~$\al{}{s}(Q^2)$ (e.g., the
one--loop Landau pole), being  artifacts of the perturbative
computations, invalidate this approach in the infrared domain. In
turn, this significantly complicates the theoretical description of
low--energy experimental data, and eventually forces one to resort to
various models and phenomenologically inspired approximations.

     The ``semi--experimental'' method for obtaining  the Adler
function in the infrared domain is to actually employ the data on the
$R$--ratio of $\epem$ annihilation into hadrons, in conjunction with
the dispersion relation~\cite{Adler}. Specifically, one first
constructs $R(s)$ by merging its low--energy experimental data with
its high--energy perturbative prediction, and then integrates the
dispersion relation for the Adler function~\cite{Adler}. This way of
modeling of the $R$--ratio entails certain complications. In
particular, one has to properly take into account the effects due to
the analytical continuation of the perturbative results for $D(Q^2)$
into the timelike domain (such as the resummation of the so--called
$\pi^2$--terms, see Ref.~\cite{RKP82}), and resolve the ambiguities
related to the matching condition between the experimental and
perturbative inputs for~$R(s)$. In addition, the $\epem$ data are not
always suitable for the extraction of the Adler function, basically
due to large systematic uncertainties in the infrared domain.
Nevertheless, recent progress in the study of the $\tau$~lepton
decays offers a way to overcome the latter difficulty. Specifically,
for the energies below the mass of the $\tau$~lepton, the $\epem$
data can be substituted (up to isospin breaking
effects~\cite{Isospin}) by a precise inclusive vector spectral
function~\cite{ALEPH}, extracted from the hadronic $\tau$~decays. It
is worthwhile to note also that some insights on the infrared
behavior of~$D(Q^2)$ may be gained from chiral perturbation theory
(see, e.g., paper~\cite{Rafael} and references therein) and lattice
simulations~\cite{Lattice}.

     An important source of the nonperturbative information about the
hadron dynamics at low energies is provided by the relevant
dispersion relations. The latter, being based on the general
principles of the local Quantum Field Theory (QFT), supply one with
the definite analytic properties in a kinematic variable of a
physical quantity at hand. The idea of employing this information
together with the perturbative treatment of the renormalization
group (RG) method forms the underlying concept of the so--called
``analytic approach'' to QFT. It was first proposed in the framework
of the  Quantum Electrodynamics and applied to the study of the invariant
charge of the theory~\cite{AQED}. Later on, it was  argued (see,
e.g., papers~\cite{Bjorken,West,Fischer,DMW,ShSol} and references
therein)  that a similar method can also be useful for studying the
non--Abelian theories. Eventually, proceeding from these motivations,
the analytic approach to Quantum Chromodynamics~(QCD) has been
developed~\cite{ShSol}. Some of the main advantages of this method
are the absence of the unphysical singularities and a fairly good
higher--loop and scheme stability of outcoming results. The analytic
approach has been successfully employed in studies of the strong
running coupling~\cite{ShSol,PRD}, perturbative series for QCD
observables (the so--called ``Analytic perturbation theory'' (APT),
see papers~\cite{APT,MS97,MSS02,MSSY} and references therein), hadron
spectrum~\cite{Prosperi}, pion form factor~\cite{Stefanis}, and some
other intrinsically nonperturbative aspects of the strong
interaction~\cite{PRD,Review}.

     The primary objective of this paper is to derive novel integral
representations for the Adler function and the $R$--ratio in a
general framework of the analytic approach to QCD. These
representations contain the nonperturbative information captured in
the dispersion relation for $D(Q^2)$, the effects due to the
interrelation between spacelike and timelike domains, and the effects
due to the nonvanishing pion mass. In addition, we compute the Adler
function within the approach at hand, by employing its perturbative
approximation as the only additional input, and compare the obtained
result with the relevant experimental data.

     The layout of the paper is as follows. In
Sec.~\ref{Sect:AdlerDisp} the dispersion relation for the Adler
function~$D(Q^2)$ and its relation to the $R$--ratio of the $\epem$
annihilation into hadrons are briefly reviewed. In
Sec.~\ref{Sect:AdlerMAPT} new integral representations for~$D(Q^2)$
and~$R(s)$, which involve a common spectral function, are derived.
The calculation of the Adler function within the developed approach,
its comparison with the experimental prediction for~$D(Q^2)$, and a
discussion of the obtained results are presented in
Sec.~\ref{Sect:Results}.

\section{Dispersion relation for the Adler function}
\label{Sect:AdlerDisp}

\begin{figure}[t]
\begin{center}
\includegraphics[width=100mm]{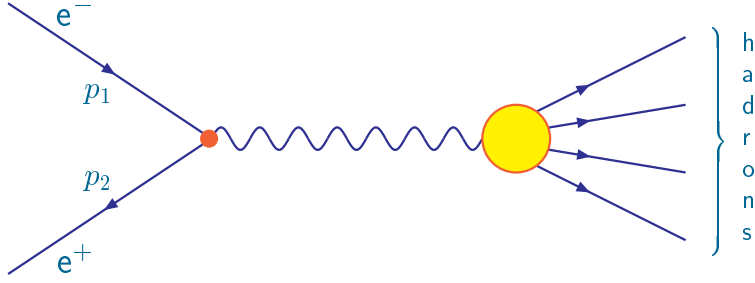}
\end{center}
\caption{The Feynman diagram for the electron--positron
annihilation into hadrons.}
\label{Plot:epem}
\end{figure}

     Let us consider the process of the electron--positron
annihilation into hadrons in the lowest order of the electromagnetic
interaction (see Fig.~\ref{Plot:epem}). The corresponding Feynman
amplitude is given by
\begin{equation}
\label{epemAmpl}
\overline{v}(p_{2}) e \gamma^{\mu} u(p_{1}) \frac{1}{q^2}
\left<\Gamma\left|J_{\mu}(q)\right|0\right>,
\end{equation}
where $u(p_{1})$ and $\overline{v}(p_{2})$ are the Dirac spinors of
the electron and positron respectively, $s=q^2=(p_{1} + p_{2})^2$ is
the timelike momentum transferred, $\Gamma$ denotes a final hadron
state,  and $J_{\mu}(q)$ stands for the quark current. The
probability of the transition of an electron--positron pair into
hadrons is proportional to the square of the
amplitude~(\ref{epemAmpl}) summed over the states~$\Gamma$. The
hadronic part of the latter can be represented as
\begin{equation}
\label{HTDef}
\Delta_{\mu\nu}(q^2)=\sum_{\Gamma}\left<0\left|J_{\mu}(-q)\right|
\Gamma\right>\left<\Gamma\left|J_{\nu}(q)\right|0\right>
= \left<0\left|J_{\mu}(-q)J_{\nu}(q)\right|0\right>,
\end{equation}
where the completeness $\sum_{\Gamma} \left|\Gamma\left>
\right<\Gamma\right|=1$ has been employed. It is worth emphasizing
here that $\Delta_{\mu\nu}(q^2)$ exists only for $q^2 \ge 4\mpi^2$,
since otherwise no hadron state $\Gamma$ could be excited. In
particular, $\Delta_{\mu\nu}(q^2)$ vanishes identically below the
two--pion threshold (i.e., for $q^2<4\mpi^2$),  see
Ref.~\cite{Feynman} for the details.

     The hadronic tensor $\Delta_{\mu\nu}(q^2)$ is proportional
to the discontinuity of the function
\begin{equation}
\label{PDef}
\Pi_{\mu\nu}(q^2)=i\int\!d^4x\,e^{i q \cdot x} \left<0\left|
T\!\left\{J_{\mu}(x)\, J_{\nu}(0)\right\}
\right|0\right> = (q_\mu q_\nu - g_{\mu\nu}q^2)\Pi(q^2)
\end{equation}
across the physical cut $q^2 \ge 4\mpi^2$ along the positive semiaxis
of real~$q^2$. In Eq.~(\ref{PDef}) $\Pi(q^2)$ stands for the hadronic
vacuum polarization function. The latter satisfies the
once--subtracted dispersion relation~\cite{Feynman,Adler}
\begin{equation}
\label{PDisp}
\Pi(q^2) = \Pi(s') - \left(q^{2}-s'\right)
\!\int_{4\mpi^2}^{\infty}\!\frac{R(s)}{(s-q^2)(s-s')}\,d s,
\end{equation}
where $\mpi=(134.9766 \pm 0.0006)\,$MeV~\cite{PDG04} is the mass of
the $\pi^{0}$ meson. In Eq.~(\ref{PDisp}) $R(s)$ denotes the
measurable ratio of two cross--sections
\begin{equation}
\label{RDef}
R(s) = \frac{1}{2 \pi i} \lim_{\varepsilon \to 0_{+}}
\left[\Pi(s - i \varepsilon) - \Pi(s + i \varepsilon)\right] =
\frac{\sigma\left(\epem \to \mbox{hadrons}; s\right)}
{\sigma\left(\epem \to \mu^{+}\mu^{-}; s\right)},
\end{equation}
with $s$ being the center--of--mass energy squared of the
annihilation process. It is worth noting here that  $R(s) \equiv 0$
for $s<4\mpi^2$, because of the kinematic restrictions mentioned
above.

     For practical purposes it proves convenient to deal with the
Adler function~\cite{Adler}, which is defined as the logarithmic
derivative of the hadronic vacuum polarization function~(\ref{PDef})
\begin{equation}
\label{AdlerDef}
D(Q^2) = \frac{d\, \Pi(-Q^2)}{d \ln Q^2},
\end{equation}
and, therefore, does not depend on the choice of subtraction
point~$s'$ in dispersion relation~(\ref{PDisp}). In
Eq.~(\ref{AdlerDef}) $Q^2 = -q^2 \ge 0$ denotes a spacelike momentum.
The Adler function~(\ref{AdlerDef}) plays an indispensable role for
the simultaneous processing of the timelike and spacelike
experimental data, see Refs.~\cite{Adler,Georgi}. Indeed, since the
perturbation theory and the RG method are not applicable directly to
the study of the observables depending on the timelike kinematic
variable, for the self--consistent description of the latter one
first has to relate the timelike experimental data with the
perturbative results. Here, the required link between the
$R$--ratio~(\ref{RDef}) and the Adler function~(\ref{AdlerDef}) can
be obtained by differentiating Eq.~(\ref{PDisp}), that results in the
dispersion relation~\cite{Adler}
\begin{equation}
\label{AdlerDisp}
D(Q^2) = Q^2 \int_{4\mpi^2}^{\infty}
\frac{R(s)}{(s + Q^2)^2} \, d s.
\end{equation}
As it has been mentioned in the Introduction, this equation is
commonly employed for restoring~$D(Q^2)$ from the~$\epem$
experimental data. On the other hand, the inverse form of
relation~(\ref{AdlerDisp}), which enables one to continue an explicit
expression for the Adler function into timelike domain, can be
derived by integrating Eq.~(\ref{AdlerDef}) between finite limits.
Ultimately, this leads to
\begin{equation}
\label{AdlerInv}
R(s) = \frac{1}{2 \pi i} \lim_{\varepsilon \to 0_{+}}
\int_{s + i \varepsilon}^{s - i \varepsilon}\!
D(-\zeta) \, \frac{d \zeta}{\zeta},
\end{equation}
where the integration contour lies in the region of analyticity of
the integrand, see Ref.~\cite{RKP82}. In particular,
Eq.~(\ref{AdlerInv}) provides the only known way for obtaining an
explicit expression for~$R(s)$, using theoretical prediction for
$D(Q^2)$ as input (see also Refs.~\cite{Adler,Georgi,Bjorken} for
details).

\section{Novel integral representations for $D(Q^2)$ and $R(s)$}
\label{Sect:AdlerMAPT}

     Before proceeding to the derivation of the integral
representations for~$D(Q^2)$ and~$R(s)$, it is worth emphasizing that
the dispersion relation~(\ref{AdlerDisp}) imposes stringent
constraints on the form of the Adler function. Specifically,
Eq.~(\ref{AdlerDisp}) implies that $D(Q^2)$ is an analytic function
in the complex $Q^2$--plane with the only cut $Q^2 \le -4\mpi^2$
along the negative semiaxis of the real~$Q^2$. In addition, given
that (i) $R(s)$, being a physical quantity, assumes finite values,
and (ii) its asymptotic ultraviolet behavior $R(s) \simeq 1 +
\mathcal{O}[\ln^{-1}(s/\Lambda^2)]$ when $s\to\infty$, one concludes
from Eq.~(\ref{AdlerDisp}) that $D(Q^2)$ vanishes in the infrared
limit~$Q^2 \to 0$.

     Let us for the moment turn off the strong interactions,
and neglect all
effects due to quark masses (these latter effects will be
disregarded throughout the paper, even when the
strong interaction are turned back on; a brief discussion
of their possible relevance will be given in the next section).
In
this case, the $R$--ratio (\ref{RDef}) is determined by the parton
model prediction~\cite{Feynman}:
\begin{equation}
\label{RPMPM}
R_{0}(s) = \theta(s - 4\mpi^2), \qquad s > 0,
\end{equation}
where the overall factor $\Nc\sum_{f}Q_f^2$ is omitted throughout,
$\Nc=3$ is the number of colors, $Q_{f}$ denotes the charge of the
quark of the $f$th flavor, and  $\theta(x)$ stands for the Heaviside
step--function. In turn, as it follows from the dispersion
relation~(\ref{AdlerDisp}), Eq.~(\ref{RPMPM}) corresponds to the
following zeroth order prediction for the Adler function:
\begin{equation}
\label{AdlerPMPM}
D_{0}(Q^2) = \frac{Q^2}{Q^2 + 4\mpi^2}, \qquad Q^2 > 0.
\end{equation}
Obviously, this expression vanishes in the infrared limit $Q^2 \to
0$, and satisfies the physical condition $D_{0}(Q^2) \to 1$ when
$Q^2\to\infty$. Thus, along with the ``standard'' intrinsically
nonperturbative contributions~\cite{SVZ,Dorokhov}, the Adler
function also receives power corrections due to the nonvanishing pion
mass, which turn out to be important for $Q \lesssim 2\,$GeV:
\begin{equation}
D_{0}(Q^2) \simeq 1 + \sum\nolimits_{n=1}^{\infty}
\left(-\frac{4\mpi^2}{Q^2}\right)^{n},
\qquad Q^2>4\mpi^2.
\end{equation}
It is worth noting that the modification of the parton model
prediction~(\ref{AdlerPMPM}) due to the mass of the $\pi$~meson is
crucial for arriving at a form of~$D(Q^2)$ which agrees with the
low--energy experimental data. Had one neglected the pion mass, one would
have instead obtained as a zeroth order approximation $D_{0}(Q^2)=1$,
which would have been inconsistent with the infrared behavior of the
Adler function extracted from the experiment (see
Fig.~\ref{Plot:Adler}). Remarkable as it may seem, to the best
of our knowledge, Eq.~(\ref{AdlerPMPM}) does not appear anywhere in
the existing literature.

\begin{figure}[t]
\begin{center}
\includegraphics[width=70mm]{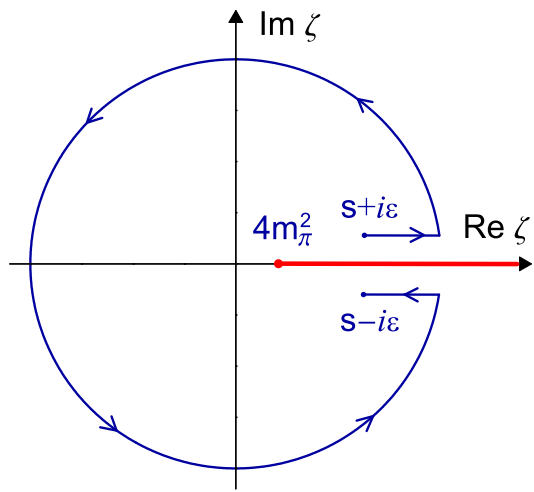}
\end{center}
\caption{The integration contour in Eq.~(\protect\ref{AdlerInv}) for
$s>4\mpi^2$. The physical cut of the Adler function $D(-\zeta)$ is
shown along  the positive semiaxis of real~$\zeta$.}
\label{Plot:Contour}
\end{figure}

     The next step is to turn the strong interactions back on, and
derive new integral  representations for~$R(s)$ and~$D(Q^2)$, which
involve a common spectral function, to be denoted by~$\rd(\sigma)$.
The inclusion of the effects due to the strong interaction  modify
Eqs.~(\ref{RPMPM}) and~(\ref{AdlerPMPM}):
\begin{eqnarray}
\label{RSC}
&&R(s) = \theta(s - 4\mpi^2) + r(s), \\
\label{DSC}
&&D(Q^2) = \frac{Q^2}{Q^2 + 4\mpi^2} + d(Q^2).
\end{eqnarray}
As it has been mentioned above, the ultraviolet behavior of the
strong correction $d(Q^2)$ can be computed in the framework of
perturbation theory, and the corresponding $r(s)$ may be obtained
from it through Eq.~(\ref{AdlerInv}). Assuming that an explicit
``exact'' expression for the Adler function~(\ref{DSC}) is available,
one can restore~$R(s)$~(\ref{RSC}) by making use of
relation~(\ref{AdlerInv}). For the energies below the two--pion
threshold (i.e., for $0 \le s < 4\mpi^2$) the integration of
Eq.~(\ref{AdlerInv}) leads to $R(s)=0$, in conformity with the
kinematic restrictions mentioned above. For the energies above the
two--pion threshold it is convenient to choose the integration
contour in Eq.~(\ref{AdlerInv}) in the form presented in
Fig.~\ref{Plot:Contour}, since the strong correction~$d(Q^2)$
vanishes at the ultraviolet asymptotic. In this case the only
nontrivial contribution into the integral along the circle of the
infinitely large radius comes from the parton model
prediction~(\ref{AdlerPMPM}), whereas the strong correction $d(Q^2)$
contributes only to the integral  along the cut of~$D(Q^2)$.
Eventually, this results in
\begin{equation}
\label{RMAPT}
R(s) = \theta(s-4\mpi^2) \left[1 + \int_{s}^{\infty}
\rd(\sigma) \frac{d \sigma}{\sigma}\right],
\end{equation}
where the spectral function is determined as the discontinuity of the
Adler function across the cut
\begin{equation}
\label{RhoDefD}
\rd(\sigma) = \frac{1}{2 \pi i} \lim_{\varepsilon \to 0_{+}}
\left[D(-\sigma - i \varepsilon) - D(-\sigma + i \varepsilon)\right],
\qquad \sigma>4\mpi^2.
\end{equation}
At the same time, $\rd(\sigma)$~(\ref{RhoDefD}) can also be related
to $R(s)$, by differentiating Eq.~(\ref{RDef}) with respect
to~$\ln s$:
\begin{equation}
\label{RhoDefR}
\rd(\sigma) = - \frac{d\,R(\sigma)}{d\,\ln\sigma},
\qquad \sigma>4\mpi^2,
\end{equation}
that allows one to extract $\rd(\sigma)$ from the relevant
experimental data. In turn, the Adler function can be represented in
terms of the spectral function $\rd(\sigma)$ by integrating the
dispersion relation~(\ref{AdlerDisp}) with $R(s)$ given by
Eq.~(\ref{RMAPT}). Carrying out the integration by parts, one arrives
at
\begin{equation}
\label{AdlerMAPT}
D(Q^2) = \frac{Q^2}{Q^2+4\mpi^2}\left[1 +
\int_{4\mpi^2}^{\infty} \rd(\sigma)\,
\frac{\sigma - 4\mpi^2}{\sigma+Q^2}\,
\frac{d \sigma}{\sigma}\right].
\end{equation}

     It is worth noting that in deriving the integral
representations~(\ref{RMAPT}) and~(\ref{AdlerMAPT}) we have employed
only  Eqs.~(\ref{AdlerDisp}) and~(\ref{AdlerInv}), the parton model
prediction~(\ref{RPMPM}), and the fact that the strong
correction~$d(Q^2)$ vanishes in the asymptotic ultraviolet limit
$Q^2\to\infty$; no additional approximations nor model--dependent
assumptions were involved. Similarly to the perturbative approach,
the QCD scale parameter~$\Lambda$ remains the basic characterizing
quantity of the theory. Besides, all the effects due to the
interrelation between the spacelike and timelike observables  are
automatically accounted for by the representations~(\ref{RMAPT})
and~(\ref{AdlerMAPT}). It is important to mention also that in the
limit of the vanishing pion mass $\mpi=0$ the derived
relations~(\ref{RMAPT}) and~(\ref{AdlerMAPT}) become identical to
those obtained within the massless APT~\cite{APT}, namely
\begin{eqnarray}
\label{RAPT}
&&R_{\mbox{\tiny APT}}(s) = 1 +
\int_{s}^{\infty}\!\rd(\sigma)\frac{d \sigma}{\sigma}, \\
\label{AdlerAPT}
&&D_{\mbox{\tiny APT}}(Q^2) = 1 +
\int_{0}^{\infty} \!\frac{\rd(\sigma)}{\sigma+Q^2} \,d \sigma.
\end{eqnarray}

\section{Discussion and conclusions}
\label{Sect:Results}

     We hasten to emphasize that the new integral representation for
the Adler function~(\ref{AdlerMAPT}) incorporates the same
nonperturbative constraints on~$D(Q^2)$ as those contained in the
dispersion relation~(\ref{AdlerDisp}); namely, $D(Q^2)$
of~(\ref{AdlerMAPT}) possesses correct analytic properties in the
$Q^2$~variable, and vanishes in the infrared limit~$Q^2 \to 0$.
However, $D(Q^2)$  of~(\ref{AdlerMAPT}) has the advantage of being
expressed as the discontinuity of itself across the
cut~(\ref{RhoDefD}).  This fact brings about certain simplifications
in the theoretical analysis of the Adler function. Specifically, as
it has been noted in the Introduction, the standard extraction
of~$D(Q^2)$ through Eq.~(\ref{AdlerDisp}) constitutes a two--step
procedure: first, one has to construct~$R(s)$ for the entire energy
range $4\mpi^2 \le s < \infty$ (encountering the complications
mentioned above), and only then integrate Eq.~(\ref{AdlerDisp}).
Instead, Eq.~(\ref{AdlerMAPT}) eliminates the intermediate step of
building up~$R(s)$, allowing the reconstruction of~$D(Q^2)$ through
the spectral function~(\ref{RhoDefD}). At the same time, one is still
able to incorporate the relevant experimental information
about~$R(s)$ into the spectral function~$\rd(\sigma)$ by virtue of
Eq.~(\ref{RhoDefR}), that, however, turns out to be a bit more
complicated technically, since the numerical differentiation of the
experimental data on~$R(s)$ is required.

     In order to exploit Eq.~(\ref{AdlerMAPT}) one should furnish an
input for the central quantity of the approach at hand,  namely, the
spectral function $\rd(\sigma)$~(\ref{RhoDefD}). In
general,~$\rd(\sigma)$ is supposed to embody all available
information about functions~$D(Q^2)$ and~$R(s)$. In this paper we
restrict ourselves to the study of the perturbative contributions
to~$\rd(\sigma)$ only.  Following the same motivations as those of
APT~\cite{APT},  the latter can be obtained by identifying~$D(Q^2)$
in Eq.~(\ref{RhoDefD}) with its perturbative approximation, which at
the $\ell$--loop level reads (see paper~\cite{RPert3L} and references
therein)
\begin{equation}
\label{AdlerPert}
D\ind{(\ell)}{pert}(Q^2) = 1 +
\sum\nolimits_{j=1}^{\ell} d_j\left[\al{(\ell)}{s}(Q^2)\right]^j,
\qquad Q^2\to\infty.
\end{equation}
It is worth noting that the resulting perturbative spectral
function~$\rp{}(\sigma)$ receives no contributions from spurious
singularities of~$\al{}{s}(Q^2)$, since the former is determined as
the discontinuity of $D\ind{}{pert}(Q^2)$ across the physical cut.
Specifically, at the one--loop level  ($d_{1}=1/\pi$) one has
$\rp{(1)}(\sigma) = (4/\beta_{0})
[\ln^{2}(\sigma/\Lambda^2)+\pi^2]^{-1}$, where $\beta_{0} = 11 - 2
\nf / 3$, and $\nf$ is the number of active quarks.

\begin{figure}[t]
\begin{center}
\includegraphics[width=100mm]{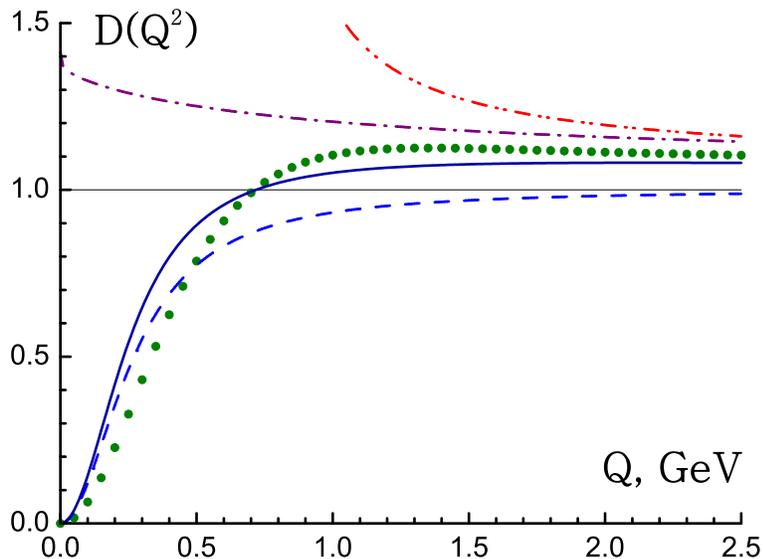}
\end{center}
\caption{Comparison of the Adler function~(\ref{AdlerMAPT})
corresponding to $\rp{(1)}(\sigma)$ (solid curve) with its
experimental behavior~($\bullet$) extracted from~\cite{ALEPH}. The
zeroth order prediction for $D(Q^2)$~(\ref{AdlerPMPM}) is  shown by
the dashed curve, the massless APT case~(\ref{AdlerAPT}) is denoted
by the dot--dashed curve, and the dot--dot--dashed curve corresponds
to the one--loop perturbative approximation~(\ref{AdlerPert}) of the
Adler function.}
\label{Plot:Adler}
\end{figure}

     It turns out that, even with the one--loop perturbative
approximation of the spectral function,~$\rp{(1)}(\sigma)$,
representation~(\ref{AdlerMAPT}) is capable of providing an output
for~$D(Q^2)$  which is compatible with its experimental prediction.
Indeed, as can be seen in Fig.~\ref{Plot:Adler},  for the entire
energy range, $0 \le Q^2 < \infty$,  the result obtained is in good
agreement with the experimental behavior of~$D(Q^2)$ extracted from
the ALEPH data on the inclusive vector spectral
function~\cite{ALEPH}, in the way described above. The value of the
scale parameter $\Lambda=335\,$MeV has been estimated  for the case
of $\nf=2$ active flavors by fitting $D(Q^2)$~(\ref{AdlerMAPT}) to
its experimental prediction by means of the least--squares method. It
is worth mentioning that the Adler function~(\ref{AdlerMAPT}) is
stable with respect to the higher--loop perturbative corrections.
Indeed, calculations up to four loops have revealed that the relative
difference between the $\ell$--loop and the $(\ell+1)$--loop
expressions for $D(Q^2)$~(\ref{AdlerMAPT}) is less than $4.9\pc$,
$1.5\pc$, and $0.3\pc$, for $\ell=1$, $\ell=2$, and $\ell=3$,
respectively, for  $0 \le Q^2 < \infty$ (the
estimation~\cite{KatStar} of the four--loop expansion coefficient
$d_{4}$ has been adopted here).

     In  order to  illustrate  the  significance of  the  mass of  the
$\pi$~meson  within   this  approach,  it  is   worth  presenting  the
massless~APT  prediction   of  $D(Q^2)$~(\ref{AdlerAPT})  computed  by
making    use     of~$\rp{(1)}(\sigma)$    (dot--dashed    curve    in
Fig.~\ref{Plot:Adler}). In  this case, one arrives at  a result, which
is free  of infrared unphysical  singularities (to be  contrasted with
the perturbative dot--dot--dashed curve in Fig.~\ref{Plot:Adler}), but
fails to describe  the experimental behavior of the  Adler function in
the low--energy domain  $Q \lesssim 1\,$GeV, where the  effects due to
the pion  mass become appreciable.   Of course, in the  framework of
the massless APT,  the infrared behavior of the  Adler function can be
greatly improved following  the procedure introduced in \cite{MSS01};
this consists  essentially in carrying out  an appropriate resummation
of  threshold  singularities, and  introducing  into (\ref{RAPT})  and
(\ref{AdlerAPT})  effects from  nonperturbative  light quark  masses.
The  necessary nonperturbative  information  on the  quark masses  is
furnished  from  the  study  of Schwinger-Dyson  equations  and  quark
condensates.     Specifically,    one     extracts     a    particular
momentum-dependence  for the dynamical  (effective) quark  mass, which
interpolates  between the  constituent  and current  mass values.   We
believe  however  that  the  method developed  here  presents  certain
advantages compared  to that of \cite{MSS01}. This is so
because,  in addition to
the  possible ambiguities  stemming  from the  truncation and  general
treatment  of  the  Schwinger-Dyson  equations,  the  final  numerical
implementation of  the procedure of \cite{MSS01} seems  to require
rather elevated  constituent  masses  for the  light
quarks  ($m_{u}=m_{d}=250\,$MeV); instead,  the  only phenomenological
parameter appearing in Eqs.~(\ref{RMAPT}) and~(\ref{AdlerMAPT}) is the
measurable  mass of  the pion.   Finally,  again in  the framework  of
massless APT,  the infrared  behavior of the  Adler function  has been
brought into qualitative agreement with experiment, but at the expense
of  resorting to additional assumptions,  such as the
vector meson dominance~\cite{Cvetic}.

As mentioned in the  previous section, in this  paper we have
omitted effects  due to the quark  masses.  Of course, this  is not to
say that such  effects are not relevant.  As a matter  of fact, and in
addition to their role in bringing the predictions of APT in agreement
with  experiment, as  exemplified in  \cite{MSS01}, they  could  be of
relevance even  within the  approach at hand.   In particular,  in the
present  analysis  such  effects  can  be retained  by  employing  the
perturbative prediction for the Adler function, which accounts for the
quark  masses  (see   paper~\cite{AdlerQM}  and  references  therein),
instead  of $D\ind{}{pert}(Q^2)$  given by  Eq.~(\ref{AdlerPert}).  We
hope to present a detailed study of this issue in the near future.

     The results obtained in this paper can be further scrutinized
through the incorporation of higher order perturbative contributions
into the spectral function~$\rd(\sigma)$, along with the effects due
to the quark masses, power corrections due to gluon and quark
condensates~\cite{SVZ,Dorokhov}, and restrictions imposed by the
low--energy experimental data. It would also be of apparent interest
to apply the present approach to the estimation of the hadronic
contributions to the muon anomalous magnetic moment and to the shift
of the electromagnetic fine structure constant.

\ack

The authors thank M.~Davier, A.E.~Dorokhov, S.\ Kluth, P.~Rakow, and
I.L.~Solovtsov for stimulating discussions and useful comments. This
work was supported by grants SB2003-0065, CICYT FPA2005-0178, and
RFBR 05-01-00992.

\end{document}